\documentclass[12pt]{article}

\usepackage{amsfonts,amssymb}
\usepackage{amsmath,emlines}

\def\tr{\mbox{tr}\,}

\def\openone{\leavevmode\hbox{\small1\kern-3.3pt\normalsize1}}

\begin{document}
\baselineskip=16pt
\begin{center}
{\bf DYNAMICAL MODELS OF ADIABATIC $N $-SOLITON INTERACTIONS}
\end{center}

\begin{center}
{\bf V. S. Gerdjikov}
\end{center}

\begin{center}
{\bf \it  Institute for Nuclear Research and Nuclear Energy, Bulgarian
Academy of Sciences,\\ 72 Tzarigradsko chaussee blvd., 1784 Sofia,
Bulgaria}
\end{center}

{\bf Abstract.}
The adiabatic $N$-soliton  train interactions for the scalar nonlinear
Schr\"o\-din\-ger (NLS) equation and its perturbed versions are well
studied.  Here we briefly outline how they can be generalized for the
higher NLS-type equations and for the multicomponent NLS equations. It is
shown that in all these cases the complex Toda chain plays fundamental
role.

\bigskip

\bigskip

{\bf 1. Introduction.}

In a number of applications to fiber optics communications (see e.g.
\cite{Agra,Arn,VSG} and the numerous references therein)  it is
important to analyze the $N $-soliton train interaction of the (perturbed)
nonlinear Schr\"odinger equation ((p)NLS) and some of its multicomponent
versions. An important class of these equations are known to be integrable
by applying the inverse scattering method to the generalized
Zakharov-Shabat system \cite{Mana}:
\begin{eqnarray}\label{eq:46.7}
L\psi \equiv \left( i {d  \over dx } + q(x,t) - \lambda J
\right) \psi (x,t,\lambda ) =0, \quad
J= \left( \begin{array}{cc} \openone & 0 \\ 0 & -\openone\end{array}
\right), \quad  q(x,t)= \left( \begin{array}{cc} 0 & \vec{u}^\dag \\
\vec{u} & 0 \end{array} \right),
\end{eqnarray}
where $q(x,t) $ and $J $ are matrices with compatible block structure. The
case when $\vec{u} $ is a $ P$-component vector is relevant to the
vector NLS known also as the Manakov model \cite{Mana}; more general
multicomponent NLS equations are obtained if $u(x,t) $ is a generic
rectangular matrix.

The class of higher multicomponent pNLS equations may be written as
\cite{PPK}:
\begin{eqnarray}\label{eq:46.3}
iJq_t + 2 f(\Lambda )q(x,t) = iJ {\cal  R}[u], \qquad
{\cal R}[u]= \left( \begin{array}{cc} 0 & R^\dag [\vec{u}] \\
R[\vec{u}] & 0 \end{array} \right),
\end{eqnarray}
where $f(\lambda ) $ is the dispersion law (for the NLS eq. $f(\lambda
)=\lambda ^2 $), and $\Lambda  $ is the recursion operator
\begin{eqnarray}\label{eq:46.5}
\Lambda =\Lambda _++\Lambda _-, \qquad
\Lambda _\pm X = {i \over 4} \left[J, {dX  \over dx } \right] - iq(x)
\int_{\pm\infty }^{x} dy\, {1  \over 2 } \tr \left( \left[J, q],
X(y)\right] \right).
\end{eqnarray}
In what follows we consider only dispersion laws $f(\lambda )=\lambda
^2g(\lambda ) $ with polynomial $g(\lambda ) $.

\medskip
{\bf 2. $N $-soliton trains of the higher scalar NLS equations.}

Let us start with a brief analysis of the $N $-soliton train interactions
for the higher scalar (i.e., $P=1 $) NLS equations; by $N $-soliton train
we mean a solution of the equation satisfying the initial condition
\begin{eqnarray}\label{eq:Nst}
u(x,0) = \sum_{k=1}^{N} u_k^{\rm 1s}(x,0)
\end{eqnarray}
where $u_k^{\rm 1s}(x,t) $ is the one-soliton solution of the higher NLS
eq. Below following the Karpman-Solov'ev parametrization \cite{KS} we
write:
\begin{eqnarray}\label{eq:46.1}
u_k^{\rm 1s}(x,t) &=& {2\nu _k e^{i\phi _k(x,t)}\over \cosh (z_k(x,t))},
\qquad z_k(x,t)=2\nu _k (x-\xi_k(t)), \qquad
\xi_k(t) = {f_{1,k}  \over \nu _k } t + \xi _{k,0}, \nonumber\\
\phi _k(x,t) &=& {2\mu _k  \over \nu _k } z_k(x,t) + \delta _k(t),
\qquad \delta _k(t) = {2 (\mu _k f_{1,k} - \nu _k f_{0,k}) \over \nu _k }
t + \delta _{k,0},
\end{eqnarray}
where $\xi_k(t) $ characterizes the center of mass position of the $k $-th
soliton and $\nu _k $, $\mu _k $ and $\delta _k $ characterize its
amplitude, velocity and phase. Here $f(\lambda _k^\pm) = f_{0,k} \pm
if_{1,k} $ is the value of the dispersion law for $\lambda _k^\pm= \mu
_k\pm i\nu _k $. As it is well known, $L $ (\ref{eq:46.7}) remains
isospectral if $q(x,t) $ satisfies (\ref{eq:46.3}) with ${\cal R}=0 $.

The adiabatic approximation means that the solitons are well separated and
have nearly equal amplitudes and velocities. If we denote by $\epsilon$
the overlap of two neighboring solitons and assume that at $t=0 $ we
have $\xi_k(0)=\xi_{k,0} $ and $\xi_{k+1,0}-\xi_{k,0} \simeq r_0 $ then
\begin{eqnarray}\label{eq:46.2}
\left| \lambda _k^+ - \lambda _0 ^+ \right|^2 \simeq \epsilon \ll 1 ,
\qquad \epsilon \simeq 16\nu _0^2 r_0 e^{-2\nu _0r_0},
\qquad \nu _0r_0 \gg 1, \qquad |\nu _0-\nu _{k,0}|r_0 \ll 1.
\end{eqnarray}
where $\lambda _0^+=\mu _0 + i\nu _0 =\sum_{k=1}^{N} (\mu _k + i\nu _k)/N
$ and $\lambda _0^-=(\lambda _0^+)^* $.

In the adiabatic approximation all $N $ solitons  keep their identity;
the main part of the energy of the train is related to the $2N $ discrete
eigenvalues of $L $ which coincide with $\lambda _k^\pm $ only in the
limit $r_0\to\infty $.  Since the solitons are well separated we can
describe the slow evolution of their parameters by deriving a dynamical
system for them.  Thus in \cite{VSG} we derived the generalized
Karpman-Solov'ev system (GKS) describing soliton trains with $N>2 $.
After some additional approximations using (\ref{eq:46.1}) the GKS
simplifies to the complex Toda chain (CTC) with $N $ sites:
\begin{eqnarray}\label{eq:ctc}
{d^2 Q_k  \over dt^2 } = C \left( e^{Q_{k+1} -Q_{k}} - e^{Q_{k} -Q_{k-1}}
\right) + {\cal  R}[Q],
\end{eqnarray}
where $C=4 g_0h_0 $, $f_0=f(\lambda _0^+) $, $g_0=g(\lambda _0^+) $, $h_0
=f(\lambda _0^+)-f(\lambda _0^-) - (\lambda _0^+-\lambda _0^-)f'(\lambda
_0) $ and
\begin{eqnarray}\label{eq:q-k}
Q_k = - 2\nu _0 \xi_k + i (2\mu _0 \xi_k  -\delta _k) + Q(t)+ ik\pi ,
\qquad {dQ  \over dt }= 2if_0 + {\lambda _0^+  \over \nu _0 } h_0.
\end{eqnarray}
${\cal  R}[Q] $ is determined by the perturbative terms ${\cal R}[u]$
in the pNLS equation, see \cite{VSG}. For $f(\lambda)=\lambda ^2 $ we
have $g_0=1 $, $C= 16\nu _0^2 $; besides the linear function $Q(t) $ can
be replaced by $1/N \sum_{k=1}^{N} \delta _k $ and thus we reproduce the
results for the $N $-soliton trains of the scalar pNLS, see
\cite{Arn,VSG,Mana}.

Note that $Q(t) $ can be adjusted as convenient. Indeed, it is a linear
function of $t $ so it does nor show up in the l.h.side of (\ref{eq:ctc}).
At the same time it does not depend on $k $ so it does not show up in the
exponentials in the r.h.side of (\ref{eq:ctc}) as well.

The importance of the CTC is based on the following facts:

i)~for ${\cal R}[u]=0$   we have ${\cal R}[Q]=0$  and we deal with a
completely integrable system. As a result, given the initial values of the
soliton parameters, we are able to predict the asymptotic regime of the
corresponding $N $-soliton train, \cite{VSG};

ii)~the CTC, in contrast with the well known real Toda chain, possesses
much richer sets of asymptotical regimes. These include in addition bound
states, mixed states, singular regimes etc. In \cite{VSG} we singled out
a special subclass of $N $-soliton bound states in which the solitons move
quasi-equidistantly.

iii)~the CTC is an universal model. Indeed, it describes the $N $-soliton
train for all the higher NLS equations. The only dependence on the
dispersion law in (\ref{eq:ctc}) is in the constant $C $ which can easily
be taken away by changing $q_k \to q_k +k\ln C $.

iv)~If the coefficients in $R[u] $ are of the order of $\epsilon  $ then
the corresponding ${\cal  R}[Q] $ becomes a constant of the order of
$\epsilon  $. As a result the perturbation drives only the center of mass
and the total phase of the $N $-soliton train and does not influence the
relative coordinates and phases, see \cite{GU}.
Our final remark here is that there are indications, showing that the CTC
may describe $N $-soliton trains also for the NLEE related to more
complicated Lax operators, e.g.  to ones depending quadratically on
$\lambda  $, \cite{PC}.

\medskip
{\bf 3. $N $-soliton trains of MNLS equations.}

Our main idea here is to show that some of the results in Section 2 can be
generalized also to a special type of MNLS equations of the form:
\begin{eqnarray}\label{eq:48.1}
i {du_p  \over  dt} + {1  \over 2 } {d^2 u_p  \over dx^2 } +
\sum_{s=1}^{P} A_{ps} |u_s|^2 u_p =iR_p[\vec{u}],
\end{eqnarray}
where $A_{ps} $ is a $P\times P $ symmetrical matrix. Some of these
multicomponent equations with $R[\vec{u}]=0 $ are integrable. They can be
written in the form (\ref{eq:46.3}) and treated by the inverse scattering
problem method applied to (\ref{eq:46.7}). Among them is the Manakov model
\cite{Mana} with $A_{ps} =(-1)^{\rho _s} $, where $\rho _s $ takes values
$0 $ and $1 $.  In these cases the soliton solutions and their
interactions in the generic case are well known \cite{Mana,PPK,JYa}.

In what follows we consider somewhat more general case with  $A_{sp} $
satisfying the condition $\sum_{s=1}^{P} A_{sp}=a^2=\mbox{const} $ for all
values of $p $. They are Hamiltonian models with
\begin{equation}\label{eq:Hmnls}
H={1 \over 2} \int_{-\infty }^{\infty } dx \left( \left(
{\partial \vec{u}^\dag \over \partial x}, {\partial \vec{u} \over \partial
x}\right) - \sum_{s, p=1}^{P} A_{sp} |u_s|^2 |u_p|^2 \right).
\end{equation}
A particular case of such MNLS with $P=2 $ and $A=\left(\begin{array}{cc}
1 & \beta \\ \beta  & 1 \end{array}\right) $ with $\beta \neq 1 $ plays
important role in nonlinear optics: it describes propagation of light in
birefringent fibers, see e.g. \cite{Agra,JYa}.
Although for such generic $A $ the MNLS (\ref{eq:48.1}) is not integrable
it allows a special type of solutions whose behavior is very close to the
ones of the scalar NLS equations.

Indeed, let us consider an $N $-soliton train of (\ref{eq:48.1}) defined
as the solution to (\ref{eq:48.1}) satisfying the initial conditions
\begin{eqnarray}\label{eq:49.1}
\vec{u}(x,0) = \sum_{k=1}^{N} \vec{n}^{(k)} u_k^{\rm 1s}(x,0),
\end{eqnarray}
where $u_k^{\rm 1s}(x,t) $ is the one-soliton solution of the scalar ($P=1
$) NLS equation and $\vec{n}^{(k)} $ is the polarization vector of the $k
$-th solitons in the train.  If we put $(\vec{n}^{(k)})_s = e^{i\varphi
_{ks}} a^{-1} $ then obviously $\sum_{s=1}^{P}A_{sp}|\vec{n}^{(k)}_s|^2
=1$ and the system (\ref{eq:48.1}) reduces to the scalar NLS equation.

Now it remains to take into account that in the  scalar case the
$N$-soliton interaction is described by the CTC. The important difference
as to the scalar case is that in the definition of the corresponding $q_k$
besides the phases $\varphi _k $ there enter also the scalar products of
the corresponding polarization vectors \cite{Thes}:
\begin{eqnarray}\label{eq:}
Q_{k+1}-Q_{k} = 2i\lambda _0^+ (\xi_{k+1}-\xi_k) + i\left(\pi -\delta
_{k+1}+\delta _k + \ln {(\vec{n}_k,\vec{n}_{k+1})  \over a^2 }\right).
\end{eqnarray}
As it is the CTC for the center of mass coordinates $q_k $ is not
sufficient to determine the dynamics for the MNLS $N $-soliton train. The
complete treatment should include also an additional system of equations
describing the dynamics of the polarization vectors $\vec{n}_k $.
Work on this is in progress.

\medskip
{\bf 4. Conclusions.}

The dynamics of the (perturbed) scalar higher NLS equations is described
by the (perturbed) CTC. The only thing that depends on the dispersion law
is the coefficient $C $ in (\ref{eq:ctc}). If the perturbations
coefficients in $R[u] $ are of the order of $\epsilon  $ then ${\cal
R}[Q] $ is a constant of the order of $\epsilon  $ whose effect is to
drive only the center of mass motion of the $N $-soliton train and the
total phase. The relative positions $\xi_{k+1}(t)-\xi_k(t) $ and the phase
differences $\delta _{k+1}(t)-\delta _k(t) $ are not influenced by these
perturbations.

In the multicomponent case we showed that some special types of
$N$-soliton trains of the MNLS equations reduce to an extension of the CTC
by additional system of dynamic equations responsible for the dynamics of
the polarization vectors $\vec{n}_k $.

\medskip
{\bf Acknowledgments.}

I have the pleasure to thank professors V. I. Karpman, E. Doktorov and V.
Shcheshnovich for useful discussions.

\end{document}